\documentclass[aps,twocolumn,showpacs,preprintnumbers,amsmath,amssymb]{revtex4}

\usepackage{graphicx}
\usepackage{dcolumn}
\usepackage{bm}

\begin{document}

\preprint{ }

\title{Thin-shell wormholes from  black holes with dilaton and monopole fields}

\author{F. Rahaman}
 \email{farook\_rahaman@yahoo.com}
\affiliation{Department of Mathematics, Jadavpur University,
Kolkata - 700 032, West Bengal, India}

\author{  A. Banerjee}
\email{ayan\_7575@yahoo.co.in} \affiliation{Dept. of Mathematics,
Adamas Institute of Technology, Barasat, North 24 Parganas -
700126, India}

\date{\today}

\begin{abstract}\noindent
We provide a new type of thin-shell wormhole  from the  black
holes with dilaton  and monopole fields. The dilaton and monopole
that built the black holes may supply fuel to construct the
wormholes. Several characteristics of this thin-shell wormhole
have been discussed. Finally, we discuss the stability of the
thin-shell wormholes with a "phantom-like" equation of state for
the exotic matter at the throat.
\end{abstract}


\maketitle

\section{Introduction}
\noindent In a pioneering work, Morris and Thorne \cite{MT1988}
have shown that wormholes are the solutions of the Einstein field
equations and are supported by exotic matter that violates the
null energy condition. It is a topological feature of spacetime
that connects widely separated regions by a throat that allows to
travel from one region to the other. Since, it is not possible to
get wormhole like geometry with normal matter in Einstein theory,
 several alternative theories, such as Brans-Dicke theory, Brain world,  C-field theory,
  Kalb-Ramond, Einstein-Maxwell theory etc.   are studied time to time
 \cite{Nandi1998,Cam2003,Lobo2007c,Rahaman2004,Rahaman2006a,Rahaman2009a,Rahaman2009b}.
Since matter source plays the crucial role for constructing
wormholes, several proposals have been proposed in literature
\cite{sus2005,lobo2005d,kuh1999,zas2005,rahaman2006f,rahaman2007d,rahaman2007e,lobo2005b,das2005,rahaman2006d,jamil2010,rahaman2009d,kuh2010}

  Visser \cite{Visser1989}   proposed a method to construct  Wormholes by surgically
grafting two black hole spacetimes together in such a way that no
event horizon is permitted to form. Usually here, wormholes are
generated from exotic three-dimensional thin shell.

Visser's approach was adopted by various authors as it is the most
simple to construct theoretically, and perhaps also practically
because it minimizes the amount of exotic matter required
\cite{Poisson1995,Lobo2003,Lobo2004,Eiroa2004a,Eiroa2004b,Eiroa2005,
Thibeault2005,Lobo2005,Rahaman2006,
Eiroa2007,Rahaman2007a,Rahaman2007b,Rahaman2007c,
Lemos2007,Richarte2008,Rahaman2008a,lemos2003,Rahaman2008b,Eiroa2008a,Eiroa2008b,Rahaman2010a,
Rahaman2010b, Rahaman2010c,Dias2010,Peter2010}.

More recently, Kyriakopoulos \cite{Kyr2006} discovered a new
black-hole solution from an action that besides gravity contains a
dilaton field and a pure ( magnetic)  monopole field. These
solutions are characterized  by three free parameters namely, the
dilaton field, the monopole charge and the ADM mass.

He considered a generalized action  as
\begin{equation}
I   =   \int  d^4x \sqrt{-g}\left[ R - \frac{1}{2}\partial_\mu\psi
\partial^\mu\psi -f(\psi)F_{\mu\nu}F^{\mu\nu} \right],
\end{equation}
where
\begin{equation} f(\psi) = g_1e^{(c+\sqrt{c^2+1})\psi} + g_2e^{(c-\sqrt{c^2+1})\psi}\end{equation}
 with c, $g_1$ and $g_2$ are real constants and R the Ricci scalar, $\psi$ representing a dilaton
 field and $F_{\mu\nu}$ corresponding to a pure ( magnetic )
 monopole field described by
\begin{equation} F = Q\sin\theta d\theta \wedge d\varphi \end{equation}
where Q is the magnetic charge. Above action readily gives the
following equations on motion:
\begin{equation} (\partial^\alpha \psi );\alpha - \frac{df}{d\psi}F_{\mu\nu}F^{\mu\nu} = 0 \end{equation}
\begin{equation} (fF^{\mu\nu});\mu = 0 \end{equation}
\begin{equation} R_{\mu \nu}=\frac{1}{2}\partial_\mu \psi \partial_\nu \psi   +2f\left(F_{\mu \sigma}F_\nu ^\sigma -\frac{1}{4}
g_{\mu\nu}  F_{\rho \sigma}F^{\rho \sigma}\right)  \end{equation}
After some straight forward calculations, Kyriakopoulos
\cite{Kyr2006} obtained the following expressions in terms of the
integration constants A,B, $\alpha$ and $\psi_0$ and generalized
black hole solutions:
\[ g_1=\frac{AB}{2Q^2}e^{-\psi_0} ,~  g_2=\frac{(\alpha-A)(\alpha-B)}{2Q^2}e^{\psi_0},~e^{\psi}
=e^{\psi_0} \left(1+\frac{\alpha}{r}\right) \]
\begin{equation}\end{equation}
where $\psi_0$ is the asymptotic value of $\psi$.

\begin{equation}\label{E:line1}
ds^2 = -f(r) dt^2 + f(r)^{-1}dr^2 + h(r) (d\theta^2+\sin^2\theta
d\phi^2),
\end{equation}
where
\begin{equation}
f(r) = \frac{(r+A)(r+B)}{r(r+\alpha)}\left(
\frac{r}{r+\alpha}\right)^{\frac{c}{\sqrt{c^2+1}}},
\end{equation}
and
\begin{equation}
h(r) = r(r+\alpha)\left(
\frac{r+\alpha}{r}\right)^{\frac{c}{\sqrt{c^2+1}}}
\end{equation}
This black hole solution is asymptotically flat and has two
horizons at $ r=-A$ and $r=-B$. The Arnowitt-Deser-Misner (ADM)
mass M is given by
\begin{equation}
M  = \frac{1}{2} \left[\alpha \left( 1 +
\frac{c}{\sqrt{c^2+1}}\right)-(A+B)\right]
\end{equation}

In this paper,  we present a new kind of thin-shell wormhole
employing such a class of black holes by means of the
cut-and-paste technique \cite{Poisson1995}. The dilaton and
monopole that built the black holes may supply fuel to construct
the wormholes.

Various aspects of this thin-shell wormhole are analyzed,
particularly the equation of state relating pressure and density.
Also it has been discussed  the attractive or repulsive nature of
the wormhole. Our final topic is to search whether    this
wormhole is stable or not.

\section{Thin-shell wormhole construction}
\noindent

From the Kyriakopoulos black hole, we can take two copies of the
region with $ r\geq a$ :

\[ M^\pm = ( x \mid r \geq a )  \]

and paste them at the hypersurface

\[ \Sigma = \Sigma^\pm = ( x \mid r = a )  \]

Here we take $ a > Max(-A,-B) $ to avoid horizon and this new
construction produces a geodesically complete manifold $ M = M^+
\bigcup M^- $ with a matter shell at the surface $ r = a $ , where
the throat of the wormhole is located.    We shall use the
Darmois-Israel formalism to determine the surface stress at the
junction boundary.

  The induced metric on $\Sigma$ is
given by
\begin{equation}
               ds^2 =  - d\tau^2 + a(\tau)^2( d\theta^2 +
               \sin^2\theta d\phi^2),
\end{equation}
where $\tau$ is the proper time on the junction surface.  Using
the Lanczos equations
\cite{Visser1989,Poisson1995,Lobo2003,Lobo2004,Eiroa2004a,Eiroa2004b,Eiroa2005,
Thibeault2005,Lobo2005,Rahaman2006,
Eiroa2007,Rahaman2007a,Rahaman2007b,Rahaman2007c,Lemos2007,Richarte2008,Rahaman2008a,
Rahaman2008b,Eiroa2008a,Eiroa2008b}, one can obtain the surface
stress energy tensor $
S_{\phantom{i}j}^i=\text{diag}(-\sigma,p_{\theta}, p_{\phi})$,
where $\sigma$ is the surface energy density and $p_{\theta}$ and
$p_{\phi}$ are the surface pressures. The Lanczos equations now
yield \cite{Eiroa2005}
\begin{equation}\label{E:sigma1}
\sigma = - \frac{1}{4\pi }\frac{h^\prime(a)}{h(a)}\sqrt{f(a) +
\dot{a}^2}
\end{equation}
and
\begin{equation}\label{E:pressure1}
p_{\theta} = p_{\phi} = p =  \frac{1}{8\pi
}\frac{h^\prime(a)}{h(a)}\sqrt{f(a) + \dot{a}^2} + \frac{1}{8\pi
}\frac{2\ddot{a} + f^\prime(a) }{\sqrt{f(a) + \dot{a}^2}}.
\end{equation}
To understand the dynamics of the wormhole, we assume  the radius
of the throat to be a function of proper time, or $ a = a(\tau)$.
Also, overdot and prime denote, respectively, the derivatives with
respect to $\tau$ and $a$. For a static configuration of radius
$a$, we need to assume   $\dot{a} = 0 $ and $\ddot{a}= 0 $ to get
the respective values of the surface energy density and the
surface pressures which are given by

\begin{equation}\label{E:sigma2}
\sigma =    \frac{\left[\frac{c\alpha}{\sqrt{c^2+1}}-(2a+\alpha)
 \right]}{4 \pi
a(a+\alpha)}\sqrt{\frac{(a+A)(a+B)}{a(a+\alpha)}\left(
\frac{a}{a+\alpha}\right)^{\frac{c}{\sqrt{c^2+1}}}}
\end{equation}
and

\begin{multline}\label{E:pressure2} p=  \frac{1}{8\pi } {\sqrt \frac{\left(
\frac{a}{a+\alpha}\right)^{\frac{c}{\sqrt{c^2+1}}}}
{a(a+A)(a+B)(a+\alpha) }}  \left[2a+A+B  \right]
  \end{multline}

\textbf{ Case -1 :   $  c \rightarrow \infty $:}
\\
The above expressions read
\begin{equation}\label{E:sigma2}
\sigma = -  \frac{\sqrt{(a+B)(a+A )}}{2 \pi (a+\alpha)^2 }
\end{equation}
and

\begin{multline}\label{E:pressure2}p_{\theta}= p_{\phi} =p= \frac{1}{8\pi }
\frac{1}{\sqrt{(a+A)(a+B)}}\left[\frac{2a +A+B}{a+\alpha} \right]
  \end{multline}

\textbf{Case -2 :   $\alpha=B= - \frac{Q^2}{M} e^{-\psi_0} $,
$A=-2M$, c=0:}
\\
The above expressions read
\begin{equation}\label{E:sigma2}
\sigma = - \frac{1}{4\pi} \frac{\left(2a-\frac{Q^2}{M}
e^{-\psi_0}\right)}{a\left(a-\frac{Q^2}{M}
e^{-\psi_0}\right)}\sqrt{1-\frac{2M}{a}}
\end{equation}
and

\begin{multline}\label{E:pressure2}p_{\theta}= p_{\phi} =p= \frac{1}{8\pi }
\frac{1}{\sqrt{1-\frac{2M}{a}}}\left[\frac{2a-\frac{Q^2}{M}
e^{-\psi_0}-2M}{a(a-\frac{Q^2}{M} e^{-\psi_0})} \right]
  \end{multline}

Observe that the energy density $\sigma$ is negative. The pressure
$p$ may be positive, however.  This would depend on the position
of the throat and hence on the physical parameters $\alpha$, $A$,
and $B$  and c defining the wormhole. Similarly, $p + \sigma$, $
\sigma + 2p  $  $ $, and  $\sigma+3p $, obtained by using the
above equations, may also be positive under certain conditions, in
which case the strong energy condition is satisfied.

\section{The gravitational field}
\noindent We now turn our attention to the attractive or repulsive
nature of our wormhole.  To perform the analysis, we calculate the
observer's four-acceleration $a^\mu = u^\mu_{\,\,;\nu} u^\nu$,
where $u^{\nu} =  d x^{\nu}/d {\tau} =(1/\sqrt{f(r)}, 0,0,0)$.  In
view of the line element, Eq. (\ref{E:line1}), the only non-zero
component is given by

\begin{multline}\label{E:acceleration}
a^r = \Gamma^r_{tt} \left(\frac{dt}{d\tau}\right)^2 \\=
\frac{1}{2}
\frac{1}{(a^2+a\alpha)^2}\left(\frac{a}{a+\alpha}\right)^{\frac{c}{\sqrt{c^2+1}}}\left[
F a^2 -Da -H\right]
\end{multline} where,
\[F= \alpha-A-B+\frac{c\alpha}{\sqrt{c^2+1}}\]
\[ D= 2AB-\frac{c\alpha(A+B)}{\sqrt{c^2+1}} \]
and
\[ E= AB\alpha - \frac{c\alpha AB}{\sqrt{c^2+1}}\]

\textbf{ Case -1 :   $  c \rightarrow \infty $:}
\\
\\
The above expression reads
\[ F = -A-B ~~;~~ D = 2AB-\alpha(A+B)~~ ; ~~E=0 ~~\]
\\
\textbf{Case -2 :   $\alpha=B= - \frac{Q^2}{M} e^{-\psi_0} $,
$A=-2M$, c=0:}
\\
\\
The above expression reads
\[ F = 2M ~~;~~ D =  2Q^2e^{-\psi_0}~ ~; ~~E= -2MQ^4e^{-2\psi_0} ~~\]
\\
A radially moving test particle initially at rest obeys the
equation of motion
\begin{equation}\label{E:motion}
\frac{d^2r}{d\tau^2}= -\Gamma^r_{tt}\left(\frac{dt}{d\tau}
\right)^2 =-a^r.
\end{equation}
If $a^r=0$, we obtain the geodesic equation.  Moreover, a wormhole
is attractive if $a^r>0$ and repulsive if $a^r<0$. These
characteristics depend on the parameters $\alpha$, $A$, and $B$
and c, the conditions on which can be conveniently expressed in
terms of the coefficients $F$, $D$, and $E$. To avoid negative
values for $r$, let us consider only the root
$r=(D+\sqrt{D^2-4FE})/(2F)$ of the quadratic equation
$Fr^2-Dr-E=0$.  It now follows from Eq.~(\ref{E:acceleration})
that $a^r=0$ whenever
 \[ \left( r -\frac{D}{2F} \right)^2
    =  \frac{ D^2 -4FE}{4F^2} .\]
For the attractive case, $a^r>0$, the condition becomes
\[ \left(r -\frac{D}{2F} \right)^2
    >\frac{ D^2 -4FE}{4F^2}. \]
For the repulsive case, $a^r<0$, the sense of the inequality is
reversed.

\section{The total amount of exotic matter}
\noindent In this section we determine the total amount of exotic
matter for the thin-shell wormhole.  This total can be quantified
by the integral
 \cite{Eiroa2005,Thibeault2005,Lobo2005,Rahaman2006,
Eiroa2007,Rahaman2007a,Rahaman2007b}
\begin{equation}
   \Omega_{\sigma}=\int [\rho+p]\sqrt{-g}d^3x.
\end{equation}
By introducing the radial coordinate $R=r-a$, we get
\[
 \Omega_{\sigma}=\int^{2\pi}_0\int^{\pi}_0\int^{\infty}_{-\infty}
     [\rho+p]\sqrt{-g}\,dR\,d\theta\,d\phi.
\]
Since the shell is infinitely thin, it does not exert any radial
pressure.  Moreover, $\rho=\delta(R)\sigma(a)$.  So
\begin{multline}\label{E:amount}
 \Omega_{\sigma}=\int^{2\pi}_0\int^{\pi}_0\left.[\rho\sqrt{-g}]
   \right|_{r=a}d\theta\,d\phi=4\pi h(a)\sigma(a)\\
=- \left[(2a+\alpha) - \frac{c\alpha}{\sqrt{c^2+1}}\right]\times
\\
\left( \frac{a+\alpha}{a} \right)^{\frac{c}{2\sqrt{c^2+1}}}
 \sqrt{\frac{(a+A)(a+B)}{a(a+\alpha)}}
\end{multline}
\textbf{ Case -1 :   $  c \rightarrow \infty $:}
\\
\\
The above expression reads
\[   \Omega_{\sigma}= -2 \sqrt{(a+A)(a+B)}\]
\\
\textbf{Case -2 :   $\alpha=B= - \frac{Q^2}{M} e^{-\psi_0} $,
$A=-2M$, c=0:}
\\
\\
The above expression reads
\[   \Omega_{\sigma}= -\left(2a-\frac{Q^2}{M} e^{-\psi_0}\right) \sqrt{1-\frac{2M}{a}}\]
\\

 This NEC violating matter can be
reduced by taking the value of $a$ closer to $r_+ = Max (-A,-B)$,
the location of the outer event horizon. The closer $a$ is to
$r_+$, however, the closer the wormhole is to a black hole:
incoming microwave background radiation would get blueshifted to
an extremely high temperature \cite{tR93}. It is interesting to
note that    total amount of exotic matter needed to support
traversable wormhole can be reduced   with the suitable choice of
the parameters Q and $\psi_0$.   One can see that total amount of
exotic matters will be reduced with the increasing of  magnetic
charge Q as well as with decrease of the asymptotic value of the
dilaton
 field $\psi_0$.

  \begin{figure}
\begin{center}
\vspace{0.5cm}
\includegraphics[width=0.4\textwidth]{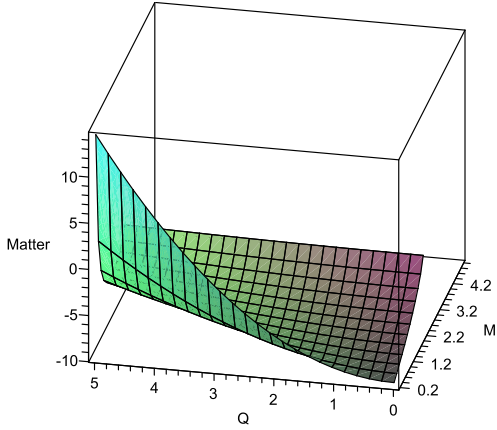}
\caption{The variation in the total amount of exotic matter on the
shell with respect to the mass ($ M $) and the charge ($ Q $) of
the black hole for   fixed values of $\psi_0 $ and throat radius
in case 2 .}
        \label{fig17}
\end{center}
\end{figure}

\section{An equation of state}
\noindent Taking the form of the equation of state (EoS) to be
$p=w\sigma$, we obtain from Eqs. (\ref{E:sigma2}) and
(\ref{E:pressure2}),
\begin{equation}\label{E:EoS}
\frac{p}{\sigma}  = w = -\frac{1}{2} \left(
\frac{a}{a+\alpha}\right)^{\frac{c}{\sqrt{c^2+1}}}
\frac{[2a^2+(A+B)(a+\alpha)+2a\alpha]}{(a+\alpha)
\left[(2a+\alpha)- \frac{c\alpha}{\sqrt{c^2+1}}\right]}.
\end{equation}

\textbf{ Case -1 :   $  c \rightarrow \infty $:}
\\
\\
The above expression reads
\[ \frac{p}{\sigma}  = w = -\frac{1}{4} \frac{(2a+A+B)(a+\alpha)}{(a+A)(a+B)}\]

\textbf{Case -2 :   $\alpha=B= - \frac{Q^2}{M} e^{-\psi_0} $,
$A=-2M$, c=0:}
\\
\\
The above expression reads
\[ \frac{p}{\sigma}  = w = -\frac{1}{2} \frac{(2a-\frac{Q^2}{M} e^{-\psi_0} -2M)}{(2a-\frac{Q^2}{M} e^{-\psi_0})}\]

Observe that if the location of the wormhole throat is very large,
i.e., if $a\rightarrow +\infty$, then $w\rightarrow -\frac{1}{2}$.
So the distribution of matter in the shell is of the
phantom-energy type.

\section{Casimir effect}

Another property worth checking is the traceless surface
stress-energy tensor $S^i_{\phantom{i}i}=0$, i.e., $ - \sigma + 2p
=0$.  The reason is that the Casimir effect with a massless field
is of the traceless type.  From this equation we find that

\begin{multline}   C \equiv    \frac{\left[2(2a+\alpha) -
\frac{c\alpha}{\sqrt{c^2+1}}\right]}{
a(a+\alpha)}\times\\\sqrt{\frac{(a+A)(a+B)}{a(a+\alpha)}\left(
\frac{a}{a+\alpha}\right)^{\frac{c}{\sqrt{c^2+1}}}}\\ +    {\sqrt
\frac{\left( \frac{a}{a+\alpha}\right)^{\frac{c}{\sqrt{c^2+1}}}}
{\frac{(a+A)(a+B)}{a(a+\alpha)} }}  \left[ \frac{\{ (\alpha
-A+B)a^2 -AB(2a+\alpha) \}}{(a^2+a\alpha)^2}  \right]= 0
\end{multline}

\textbf{ Case -1 :   $  c \rightarrow \infty $:}
\\
\\
The above expression reads
\begin{multline}   C \equiv   \frac{\sqrt{(a+B)(a+A )}}{2 \pi (a+\alpha)^2 } \\
+\frac{1}{\sqrt{(a+A)(a+B)}}\left[\frac{2a +A+B}{a+\alpha}
\right]= 0
\end{multline}

\textbf{Case -2 :   $\alpha=B= - \frac{Q^2}{M} e^{-\psi_0} $,
$A=-2M$, c=0:}
\\
\\
The above expression reads
\\
\\
\begin{multline}  C \equiv   \frac{\left(2a-\frac{Q^2}{M}
e^{-\psi_0}\right)}{a\left(a-\frac{Q^2}{M}
e^{-\psi_0}\right)}\sqrt{1-\frac{2M}{a}}
 \\+\frac{1}{\sqrt{1-\frac{2M}{a}}}\left[\frac{2a-\frac{Q^2}{M}
e^{-\psi_0}-2M}{a(a-\frac{Q^2}{M} e^{-\psi_0})} \right] =
0\end{multline}
 \begin{equation} \end{equation}

On can note that  the no real values of $a$ exist which satisfy
these equations. This  ensures that this situation could not occur
when dealing with  thin-shell wormholes. This result is rather
unfortunate as we expected Casimir effect would be  associated
with massless fields confined in the throat.

\section{Stability}
\noindent We analyze the stability taking specific equation of
state  at the throat.

We re-introduce an equation of state between the surface pressure
$p$ and surface energy density $\sigma$ as
\begin{equation} p=w\sigma   \end{equation}
 with  $  w < 0$.

This is analogous to dark energy equation of state. Here $ p$ and
$\sigma $ obey the conservation equation

\begin{equation}
               \frac {d}{d \tau} [\sigma h(a)] + p \frac{d}{d \tau}[h(a)]= 0
               \end{equation}
or
\begin{equation}
               \dot{\sigma} +    \frac{\dot{h}}{h}( p + \sigma ) = 0.
               \end{equation}
In the above equations, the overdot   denotes,   the derivative
with respect to $\tau$.

 Using the above specific equation of state, the
equation (27) yields
\begin{equation} \sigma(a) = \sigma_0 \left( \frac{h_0}{h} \right)^{ (1+w)}    \end{equation}
where, $a_0$ being initial position of the throat with $ \sigma_0
= \sigma(a_0) $ and $h_0 = h(a_0)$.

Rearranging equation (13), we obtain the thin shell's equation of
motion
\begin{equation}
\dot{a}^2 + V(a)= 0.
\end{equation}
Here the potential $V(a)$ is defined  as
\begin{equation}
V(a) =  f(a) - \left[\frac{4\pi
\sigma(a)h(a)}{h^\prime(a)}\right]^2.
\end{equation}

Now substituting the value of $\sigma(a)$ in the above equation,
we obtain the following form of potential as
\begin{equation}
V(a) = f(a) - \left[\frac{4\pi  \sigma_0 h(a)}{h^\prime(a)}\left(
\frac{h_0}{h} \right)^{(1+w)}\right]^2
\end{equation}
The explicit expression for V(a) is given by
\begin{multline}
V(a) = \frac{(a+A)(a+B)}{a(a+\alpha)}\left(
\frac{a }{a+\alpha}\right)^{\frac{c}{\sqrt{c^2+1}}} -\\
\left[\frac{L\{ a(a+\alpha)\left( \frac{a+\alpha}{a}
\right)^{\frac{c}{\sqrt{c^2+1}}} \}^{-w}}{\{(
2a+\alpha)-\frac{c\alpha}{\sqrt{c^2+1}}\}\left( \frac{a+\alpha}{a}
\right)^{\frac{c}{\sqrt{c^2+1}}}}\right]^2
\end{multline}
where $ L = 4\pi \sigma_0 h_0 ^{ (1+w)}$.
\\

\textbf{ Case -1 :   $  c \rightarrow \infty $:}
\\
The above expression reads
\begin{multline}
V(a) = \frac{(a+A)(a+B)}{(a+\alpha)^2}  - \left[\frac{L_1\
(a+\alpha)^{-2w-1}}{ 2  }\right]^2
\end{multline}

\textbf{Case -2 :   $\alpha=B= - \frac{Q^2}{M} e^{-\psi_0} $,
$A=-2M$, c=0:}
\\
\\
The above expression reads

\begin{multline}
V(a) = (1-\frac{2M}{a}) - \left[\frac{L_2 \left(a^2-
a\frac{Q^2}{M} e^{-\psi_0}\right) ^{-w}}{\left( 2a- \frac{Q^2}{M}
e^{-\psi_0}\right) }\right]^2
\end{multline}

We analysis stability by means of the figures.\\ The plot (fig 2)
indicates that $V(a)$ has a local minimum at some $a$. In other
words, it is stable in case 1. However, the plot (fig 3) indicates
that $V(a)$ has a local maximum at some $a$. In other words, it is
unstable in case 2. Thus for some specific choices of the
parameters, the thin-shell wormholes constructed from the  black
holes with dilaton  and monopole fields are stable.
\\
\begin{figure}
\begin{center}
\vspace{0.5cm}
\includegraphics[width=0.3\textwidth]{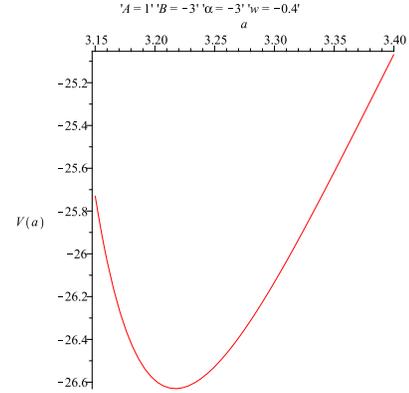}
\caption{ The variation of $V(a)$  with respect to $a$ in case 1.}
        \label{fig17}
\end{center}
\end{figure}
\begin{figure}
\begin{center}
\vspace{0.5cm}
\includegraphics[width=0.3\textwidth]{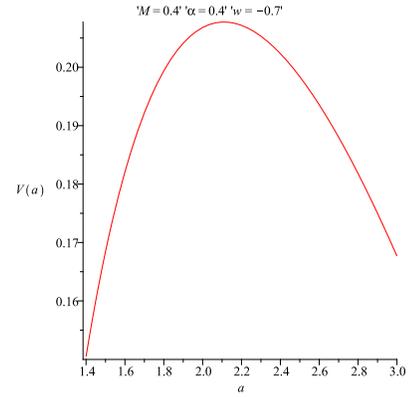}
\caption{The variation of $V(a)$  with respect to $a$ in case 2.}
        \label{fig17}
\end{center}
\end{figure}

\section{Final remarks}\noindent
An exact black hole solution with a dilaton and a pure monopole
field and its generalization were developed by  Kyriakopoulos
\cite{Kyr2006}.  Our  aim in this article is to search the new
type of thin shell wormholes from the dilaton and monopole fields
that built the black holes may supply fuel in the throat to
construct the wormholes.
 We analyzed various aspects of this
wormhole, such as the amount of exotic matter required, the
attractive or repulsive nature of the wormhole, and a possible
equation of state for the thin shell.
 Since the energy density $\sigma$ ( exotic matter )  is confined within the thin-shell,
  so later we
  demand that it obeys  some specific  form of equation of state. We have
  assumed phantom-like equation of state and this yields
explicit closed-form expression   for $\sigma$.

 We have discussed the stability of the
thin-shell wormholes with a "phantom-like" equation of state for
the exotic matter at the throat. This  approach to the stability
analysis is   different from the other method \cite{Poisson1995}of
the stability of the configuration under small perturbations
around a static solution at $a_0$. It has been shown that for some
specific choices of the parameters, the thin-shell wormholes
constructed from the black holes with dilaton  and monopole fields
are stable unlike the wormholes constructed from two Schwarzschild
spacetimes \cite{Peter2010}.

\subsection*{Acknowledgments}

 FR is thankful to Inter-University Centre for
Astronomy and Astrophysics, Pune, India for providing Visiting
Associateship under which a part of this work is carried out. FR
is also grateful to PURSE, Govt. of India
 for financial
support.

\end{document}